\documentclass[12pt,preprint]{aastex}

\slugcomment{To appear in ApJ (Letters)}

\shorttitle{Compact Centimeter and Millimeter Sources in NGC 6334 I(N)}
\shortauthors{Rodr\'\i guez et al.}

\begin{document}


\title{Compact Centimeter and Millimeter Sources \\
    in NGC 6334 I(N): OB Stars in the Making?}


\author{Luis F. Rodr\'\i guez and Luis A. Zapata\altaffilmark{1}}
\affil{Centro de Radioastronom\'\i a y Astrof\'\i sica, UNAM, Morelia 58089, M\'exico}
\email{l.rodriguez, l.zapata@astrosmo.unam.mx}

\and

\author{Paul T. P. Ho\altaffilmark{1}}
\affil{Academia Sinica Institute of Astronomy and Astrophysics, Taipei 106, Taiwan}
\email{pho@asiaa.sinica.edu.tw}


\altaffiltext{1}{Harvard-Smithsonian Center for Astrophysics,
and 
Submillimeter Array, 60 Garden Street, Cambridge, MA 02138, USA}


\begin{abstract}
We present sensitive, high angular resolution 1.3 cm and 7 mm observations of
the massive core NGC 6334 I(N), a region known to be undergoing
massive star formation. At 1.3 cm we detect three sources, of which two
had been previously detected at centimeter or millimeter wavelengths.
At 7 mm we detect four sources.
We suggest that three of these sources are subcomponents of the
millimeter source SMA 1, that at these wavelengths is the dominant source in the
region. The fourth 7 mm source appears to be associated with the 
relatively isolated source SMA 6. In all four 7 mm
sources, the continuum emission is arising from structures of
dimensions in the order of 1000 AU for which we estimate masses
of order a few solar masses. We interpret these 7 mm sources as massive
circumstellar disks that, however, surround stars or compact small stellar
groups that at present have
masses comparable to those of the disks but that may be accreting on their way to
become massive stars. 

\end{abstract}


\keywords{stars: formation --- ISM: individual (NGC 6334 I(N))}



\section{Introduction}

Located at the NE of the molecular cloud/H II region complex NGC~6334,
the massive cores NGC 6334 I and NGC 6334 I(N) are separated by only
90${''}$ and are considered ``twin'' objects
in some aspects but also very different in other characteristics
(Beuther et al. 2005). 
Both cores have similar dimensions: $\sim 10{''}$ (Sandell 2000) or
0.08 pc at a distance of 1.7 kpc (Neckel 1978).
They also have similar masses in gas ($\sim$200 M$_\odot$ for 
NGC 6334 I and $\sim$400 M$_\odot$ for 
NGC 6334 I(N); Sandell 2000), but their bolometric luminosities are very
different ($\sim 2.6 \times 10^5$ L$_\odot$ for 
NGC 6334 I and $\sim 1.9 \times 10^3$ L$_\odot$ for 
NGC 6334 I(N); Sandell 2000). 

These differences have been attributed to NGC 6334 I(N) being less
evolved than NGC 6334 I (Thorwirth et al. 2003; Hunter et al.
2006). With time, the forming massive stars in 
NGC 6334 I(N) will reach their final masses and shine with much
larger luminosity than now. There is recent observational evidence that the formation
of massive stars takes place in a similar way to that of solar type
objects, namely by accretion via a circumstellar disk: a handful of
cases where a disk could be present around a forming massive star
have been presented in the literature (e. g. Shepherd \& Kurtz 1999;
Patel et al. 2005;
Schreyer et al. 2006).
These disks are larger than those found in forming solar stars,
and have dimensions of order 1000 AU and masses of a few solar masses.
In this paper we present 1.3 cm and 7 mm observations of the core
NGC 6334 I(N), trying to study the environment of the massive stars
that are suspected to be forming there. 
 
\section{Observations}


The 1.3 cm observations were made in the BnA configuration
of the VLA of the NRAO\footnote{The National Radio 
Astronomy Observatory is operated by Associated Universities 
Inc. under cooperative agreement with the National Science Foundation.},
during 2005 January 23. The central frequency observed was
22.46 GHz and we integrated on-source for a total of
120 minutes. The absolute amplitude
calibrator was 1331+305 (with an adopted flux density of 2.52 Jy)
and the phase calibrator was 1720$-$358 (with a bootstrapped flux density
of 0.652$\pm$0.005 Jy). The phase noise rms was about 25$^\circ$,
indicating good weather conditions. The phase center of these observations was at
$\alpha(2000) = 17^h~20^m~54\rlap.^s9;~\delta(2000) = -35^\circ~44{'}~54^{''}$.
The 7 mm observations were made in the CnB configuration
of the VLA during 2005 June 30. The central frequency observed was
43.34 GHz and we integrated on-source for a total of
100 minutes. The absolute amplitude
calibrator was 1331+305 (with an adopted flux density of 1.45 Jy)
and the phase calibrator was 1720$-$358 (with a bootstrapped flux density
of 1.02$\pm$0.01 Jy). The phase noise rms was about 10$^\circ$,
indicating excellent weather conditions. The phase center of these observations was at
$\alpha(2000) = 17^h~20^m~54\rlap.^s9;~\delta(2000) = -35^\circ~45{'}~06^{''}$.

The data were acquired and reduced using the recommended VLA procedures
for high frequency data, including the fast-switching mode with a
cycle of 120 seconds. The effective bandwidth of the observations
was 100 MHz. We 
made images with natural weighting to
obtain the highest signal-to-noise possible.




\section{Analysis}

\subsection{1.3 cm Data}

The rms noise of the 1.3 cm image is 67 $\mu$Jy.
At this wavelength we detected three sources, whose positions and total flux densities
are given in Table 1. All three sources appear unresolved ($\leq 0\rlap.{''}2 -
0\rlap.{''}3$). 
The first source in Table 1 does not have a counterpart at other wavelengths
and we refer to it as VLA-K 1 (since it was detected at 1.3 cm, the so-called K band).
The other two sources are associated, respectively, with the sources
SMA 4 and SMA 1 reported by Hunter et al. (2006).
In Figure 1 we show a contour image of these last two sources as well
as the positions of other sources in the region.

\subsection{7 mm Data}

The rms noise of the 7 mm image is 320 $\mu$Jy.
At this wavelength
we detected four sources, whose positions and total flux densities
are given in Table 2. The first source in Table 2 is associated with
SMA 6, a source reported by Hunter et al. (2006). In Figure 2 we show a contour image of
this 7 mm source. The other three 7 mm sources are very close among them, forming a
line that extends by $\sim$ 4${''}$ (see Figure 3). We interpret these three sources
as the subcomponents of the 1.3 mm source SMA 1, the dominant millimeter object in the
region (Hunter et al. 2006). SMA 1 extends by about the
same extent and position angle as the three 7 mm sources.
We then refer to the three 7 mm sources as
SMA 1a, 1b, and 1c (see Table 2).  

\subsection{The Nature of the Continuum Emission}

Of the three sources detected at 1.3 cm, only one (SMA 4) has an associated 3.6 cm
counterpart, an unresolved source with flux density of 0.34$\pm$0.06 mJy 
detected by Carral et al. (2002). The other two sources detected at
1.3 cm do not have a 3.6 cm counterpart at the 4-$\sigma$ level of $\sim$0.24 mJy.
The spectral index between 3.6 cm and 1.3 cm is then 0.4$\pm$0.3 
($S_\nu \propto \nu^{0.4\pm0.3}$), consistent
with free-free emission from a thermal jet
(Reynolds 1986), although an optically-thin HII region
cannot be ruled out. This spectral index determination is reliable since
the 3.6 and 1.3 cm observations have similar angular resolution
($\sim 0\rlap.{''}2 - 0\rlap.{''}3$). The other two 1.3 cm sources do not have a
3.6 cm counterpart and lower limits of $\sim$0.9 are estimated for their spectral indices.
Again, this spectral index is consistent with the values expected for
free-free emission from a thermal jet. We suggest that the compact 1.3 cm emission detected by
us is probably tracing ionized jets.

Of the four sources detected at 7 mm, only one (SMA 1b) has 
an associated 1.3 cm
counterpart, an unresolved source with flux density of 0.57$\pm$0.12 mJy detected by us
(see Tables 1 and 2 and Figure 1). The spectral index of this source 
between 1.3 cm and 7 mm is
large, 4.5$\pm$0.4, suggesting that the 7 mm emission cannot be attributed
to optically-thick free-free emission and is most probably arising from
dust. A similar conclusion is reached for the other three 7 mm sources that are
not detected at 1.3 cm, but show a flux density of several mJy at 7 mm.
The non detection of these three sources at 1.3 cm, indicates that
their spectral index between 7 mm and 1.3 cm is $\geq$3.3.
We then suggest that, in contrast to the 1.3 cm emission that appears
to be tracing free-free emission, the 7 mm emission is tracing dust.
The 1.3 cm data has better angular resolution ($\sim0\rlap.{''}3$) than
the 7 mm data ($\sim0\rlap.{''}6$). However, we have tapered the
1.3 cm data to produce images of similar angular resolution to those
at 7 mm and our results are not changed. 

We finally compare our 7 mm data with the 1.3 mm data of Hunter et al. (2006).
This comparison is not direct since our angular resolution is of order
$0\rlap.{''}6$ while that of Hunter et al. (2006 ) is of order
$1\rlap.{''}6$. In particular, as noted before,
the source SMA 1 of Hunter et al. (2006) that is
the dominant 1.3 mm source of the region is resolved by us into three subcomponents
at 7 mm
(SMA 1a, 1b, and 1c). The total flux density of SMA 1 at 1.3 mm is 2.04 Jy.
In a solid angle similar to that occupied by SMA 1 at 1.3 mm, we detect a total
flux density of $\sim$40 mJy at 7 mm (of which 31 mJy can be attributed to the
sources SMA 1a, 1b, and 1c). A spectral index of 2.4$\pm$0.2 is derived between
7 and 1.3 mm. A similar comparison can be made for the source SMA 6, for which we
estimate
a total flux density of $\sim$9 mJy for a solid angle similar to that occupied by SMA 6 at 1.3 mm.
The same spectral index, 2.4$\pm$0.2, as derived for SMA 1 is obtained for SMA 6 between 
7 and 1.3 mm. We then suggest that at 7 and 1.3 mm we are observing 
optically thin dust emission with a dust mass opacity coefficient that goes with
frequency as $\kappa_\nu \propto \nu^{0.4}$. With this information we can estimate the
masses of the 7 mm sources. Following Hunter et al. (2006) we adopt a value of
$\kappa_{1.3mm}$ = 1.5 cm$^{2}$ g$^{-1}$
(the average of the values of 1.0 cm$^{2}$ g$^{-1}$, valid for grains
with thick dust mantles, and 2.0 cm$^{2}$ g$^{-1}$, valid for
grains without mantles). This implies $\kappa_{7mm}$ = 0.78 cm$^{2}$ g$^{-1}$. 
Assuming optically thin, isothermal dust emission and a gas-to-dust ratio of 100
(Sodroski et al. 1997), the  
total mass of the 7 mm sources is given by:

$$\Biggl[{{M} \over {M_\odot}}\Biggr] = 0.11 \Biggl[{{S_\nu} \over {mJy}}\Biggr]
\Biggl[{{T} \over {100~K}}\Biggr]^{-1}
\Biggl[{{\nu} \over {43~GHz}}\Biggr]^{-2.4}
\Biggl[{{D} \over {kpc}}\Biggr]^{2},$$

\noindent where $T$ is the dust temperature and $D$ is the distance to the source. Assuming
$T$ = 100 K for SMA 1 and  $T$ = 33 K for SMA 6 (Hunter et al. 2006), we derive masses
on the order of a few solar masses for the four 7 mm sources (see Table 2).
These dust temperatures (Hunter et al. 2006) have been estimated from molecular
observations and have uncertainties of a factor of two, that reflect in the
mass determination.

We finally discuss the different spectral indices
for the emission of the sources determined by us for the 1.3 mm to 7 mm
range (that gives a value of $\sim$2.4), and for the 7 mm to 1.3 cm
range (that gives a value of $\geq$3.3). As noted by  
Wilner et al. (2005), these changes in spectral index most probably denote 
grain growth to centimeter-size particles. 
The significant presence of centimeter-size dust grains flattens the spectral
index for wavelengths shorter than the characteristic size of the grains.

\subsection{Massive Circumstellar Disks?}

The dimensions ($\sim$1000 AU) and masses (a few $M_\odot$) of the four sources
detected at 7 mm are suggestive of massive circumstellar disks.
Furthermore, the two 7 mm sources with deconvolved dimensions (SMA 1a and SMA 1b;
see Table 2) have orientations consistent with hosting the exciting sources of
two outflows reported in the region. Megeath et al. (1999) detected an SiO outflow
whose axis is oriented at position angle of $\sim$135$^\circ$ and could be attributed 
to the source associated with SMA 1a, whose major axis is oriented at
a position angle of $\sim$42$^\circ$, nearly perpendicular to the outflow.
Hunter et al. (2006) propose that an elongated infrared nebulosity could be a jet or a
reddened reflection nebulosity from an outflow
cavity and propose that it is excited
by SMA 4. This nebulosity extends approximately to the west of the core of 
NGC 6334 I(N), with position angle of $\sim240^\circ$
and could alternatively be excited by SMA 1b, since this
source (as SMA 4) also lies along the axis of the nebulosity. 

There is, however, a problem with the interpretation of these structures
as massive circumstellar disks. In the cases of forming objects 
where the masses of the disk and its
central star have been estimated, they usually have a ratio of order
$M_{star}/M_{disk}~\simeq$ 10 (Rodr\'\i guez et al. 1998; Guilloteau \& Dutrey 1998; 
Schreyer et al. 2006).
If we apply this mass ratio to the 7 mm sources in NGC 6334 I(N), stars
with masses of tens of solar masses would be required at the center of
the disks. This is clearly not the case since the modest bolometric luminosity
of $\sim 1.9 \times 10^3$ L$_\odot$ (Sandell 2000) would then be exceeded
by one to two orders of magnitude. 

However, the ratio of $M_{star}/M_{disk}~\simeq$ 10 is not an established rule. 
For forming objects, surely the $M_{star}/M_{disk}$ ratio will
depend on the stage of evolution, with smaller values
at the initial stages.
The theoretical models of Yorke, Bodenheimer, \& Laughlin (1995)
predict values of $M_{star}/M_{disk}~\simeq$ 4 for forming intermediate-mass
stars.
It is known from theoretical two-dimensional calculations that forming stars with
$M_{star}/M_{disk}~\leq$ 3 will face gravitational instabilities
that will induce spiral waves (Laughlin \& Bodenheimer 1994).
However, from their  global, three-dimensional smoothed particle hydrodynamics simulations  
Lodato \& Rice (2005)
conclude that these massive self-gravitating disks can
survive for relatively long periods.
The problem of disk stability in massive stars has been recently reviewed
by Cesaroni et al. (2006).

From the observational point of
view, in some sources the possibility that $M_{star}/M_{disk}$ is as
low as $\sim$2 is consistent with the data (Patel et al. 2005).
Furthermore, in the case of G192.16-3.82, the protostar and disk have comparable
masses of about 8 $M_\odot$ (Shepherd, Claussen, \& Kurtz 2001). 
With the finding of additional young objects that may have comparable masses
in the disk and the protostar, it is important to review the issue of
the stability of these structures.
It is also possible that these massive disks are forming a compact stellar
group, as proposed by Sollins \& Ho (2005) for the
100 $M_\odot$ disk in G10.6$-$0.4, in which case part of the mass now observed in them
will end in the smaller members of the group. 

In summary, we believe that the observations discussed here are
consistent with the presence in NGC 6334 I(N) of several forming massive
stars that at present have reached only a fraction of their final masses
and are surrounded by disks with masses comparable to those of the stars.
Whether these massive structures that we identify as
disks are forming a single star or a compact stellar group 
cannot be disentangled with the present observations.

\acknowledgments

LFR and LAZ acknowledge the support
of CONACyT, M\'exico and DGAPA, UNAM.



{\it Facilities:} \facility{VLA}

\clearpage



\begin{figure}
\epsscale{.80}
\plotone{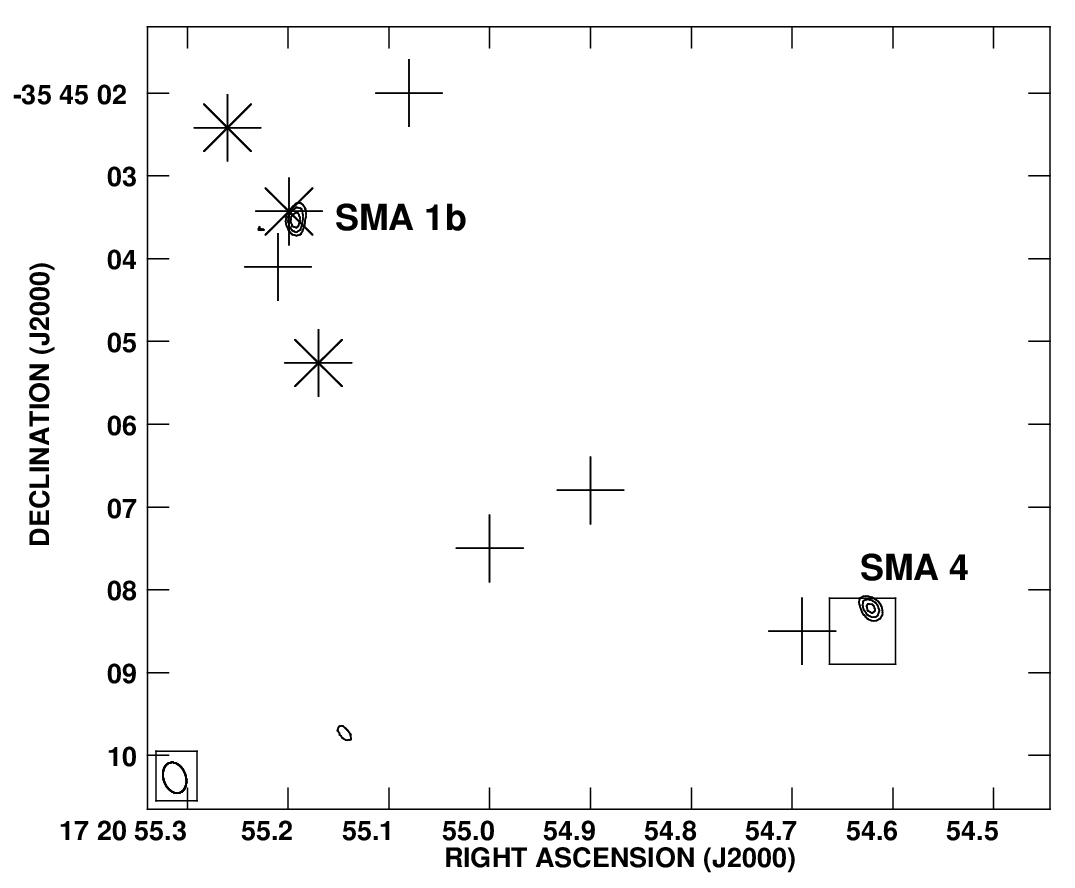}
\caption{Contour image of the 1.3 cm emission of the core
of NGC 6334 I(N). Two of the three sources detected at
this wavelength, SMA 4 and SMA 1b, are included in this region
(see Table 1 for the positions and flux densities of these
three sources). Contours are
-4, 4, 5, and 6 times 67 $\mu$Jy, the rms noise of the image. 
The half power contour of the synthesized beam ($0\rlap.{''}38 \times 0\rlap.{''}27$
with a position angle of $+19^\circ$) is shown in the bottom left corner.
The center of the empty square marks the position of one of the
two 3.6 cm sources
detected by Carral et al. (2002). The crosses mark the position and
positional error of five of the seven 1.3 mm 
sources reported by Hunter et al. (2006), while
the asterisks mark the position of three of the 
four 7 mm sources reported in this
paper.
\label{fig1}}
\end{figure}

\clearpage

\begin{figure}
\epsscale{.80}
\plotone{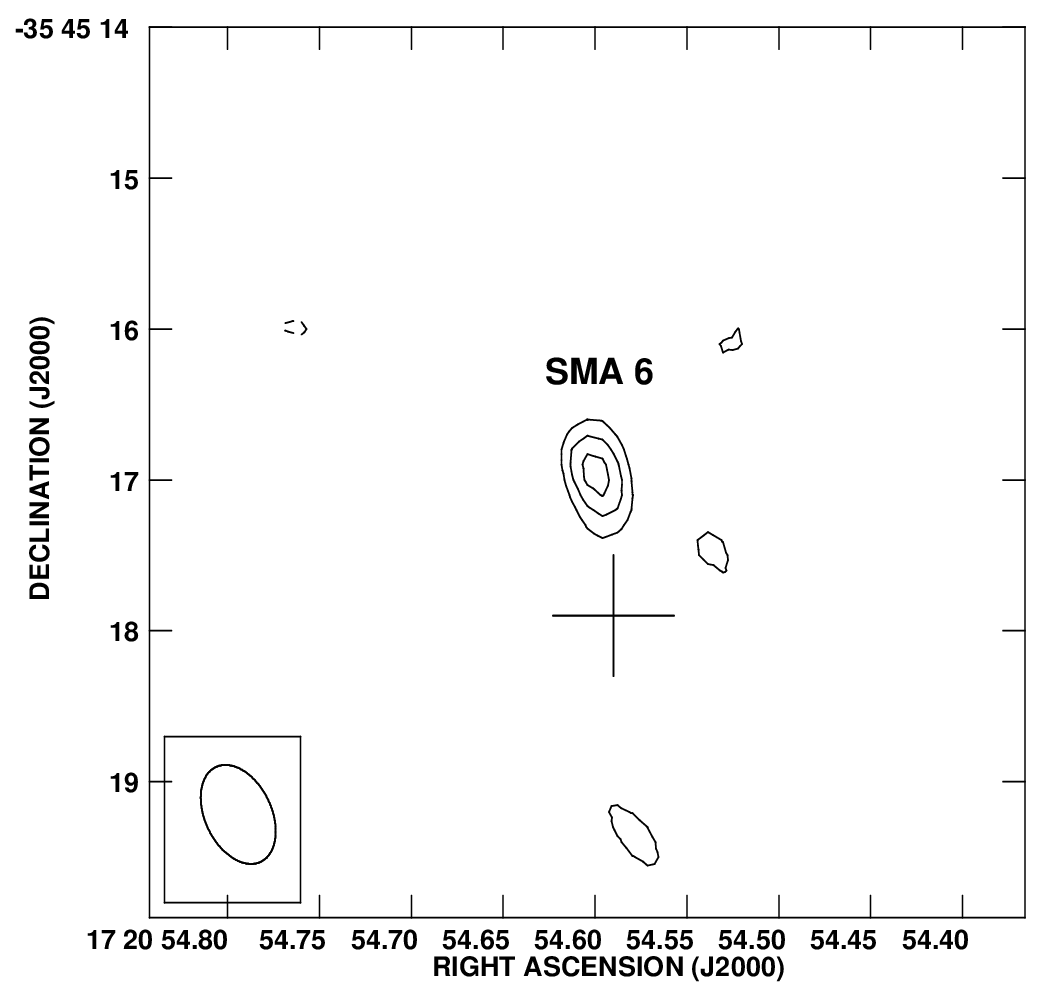}
\caption{Contour image of the 7 mm emission of the source associated
with the millimeter object SMA 6.
Contours are
-3, 3, 4, and 5 times 320 $\mu$Jy, the rms noise of the image.
The half power contour of the synthesized beam ($0\rlap.{''}69 \times 0\rlap.{''}44$
with a position angle of $+25^\circ$) is shown in the bottom left corner.
The cross marks the position and
positional error of the source SMA 6
detected by Hunter et al. (2006).
\label{fig2}}
\end{figure}

\clearpage

\begin{figure}
\epsscale{.80}
\plotone{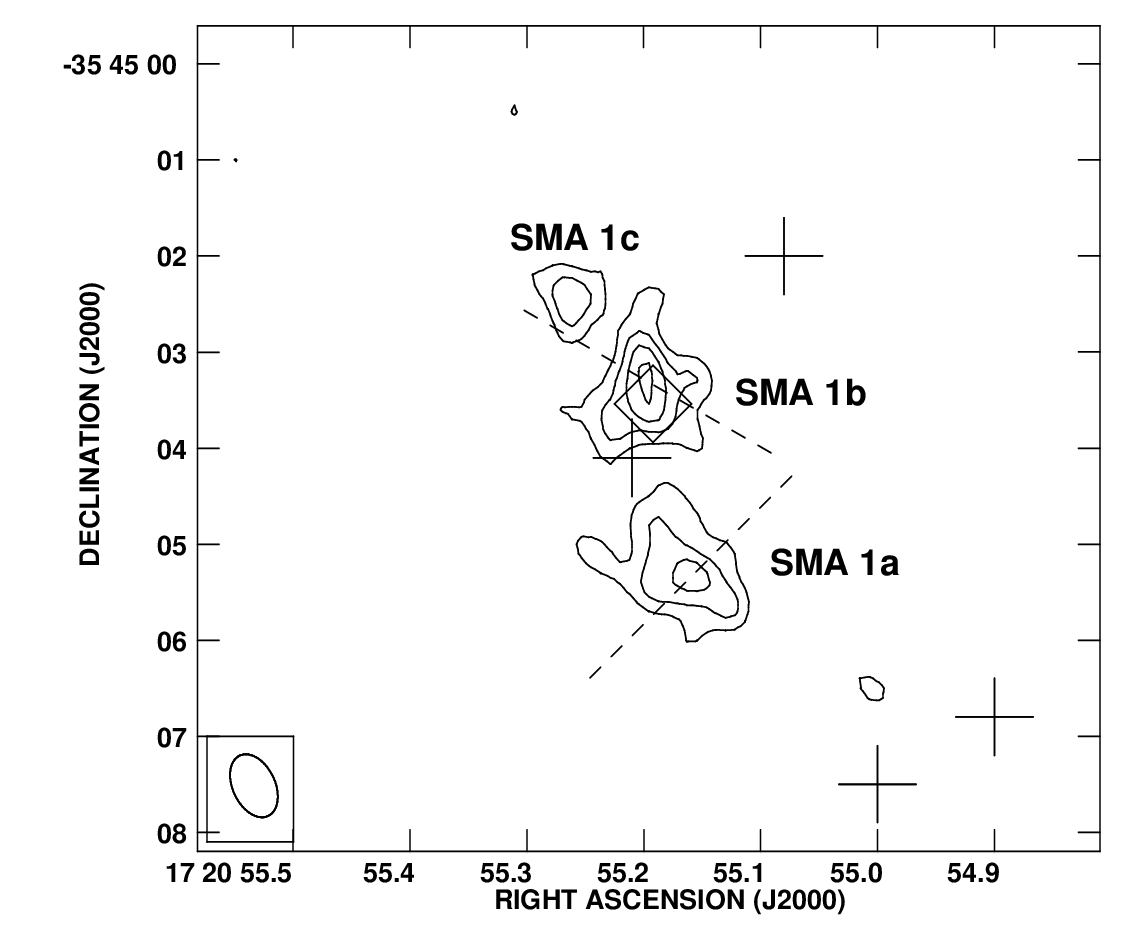}
\caption{Contour image of the 7 mm emission of the core
of NGC 6334 I(N). Three of the four sources detected at
this wavelength, SMA 1a, SMA 1b, and SMA 1c, are included in this region
(see Table 2 for the positions and flux densities of these
four sources). Contours are
-3, 3, 4, 5, and 6 times 320 $\mu$Jy, the rms noise of the image.
The half power contour of the synthesized beam ($0\rlap.{''}69 \times 0\rlap.{''}44$
with a position angle of $+25^\circ$) is shown in the bottom left corner.
The dashed lines show the orientations of two outflows in the region that could be
associated with the sources SMA 1a and SMA 1b. 
The center of the empty diamond marks the position of one of the
three 1.3 cm sources
reported in this paper (see Fig. 1 and Table 1). The crosses mark the position and
positional error of four of the seven 1.3 mm
sources reported by Hunter et al. (2006).
\label{fig3}}
\end{figure}








\clearpage

\begin{table}
\begin{center}
\caption{Parameters of the 1.3 cm Sources in NGC 6334 I(N).\label{tbl-2}}
\begin{tabular}{lccc}
\tableline\tableline
 &\multicolumn{2}{c}{Position$^a$} & Total Flux  \\
\cline{2-3} 
Source &  $\alpha$(J2000) & $\delta$(J2000) & Density$^b$ (mJy) \\
\tableline
VLA-K 1 & 17 20 52.701 & -35 45 13.52 & 0.59$\pm$0.09 \\ 
SMA 4 & 17 20 54.622 & -35 45 08.23 & 0.50$\pm$0.11 \\
SMA 1b & 17 20 55.192 & -35 45 03.54 & 0.57$\pm$0.12 \\
\tableline
\end{tabular}
\tablenotetext{a}{Units of right 
ascension are hours, minutes, and seconds
and units of declination are degrees, arcminutes, and arcseconds. Positional accuracy
is estimated to be $0\rlap.{''}05$.}
\tablenotetext{b}{All three sources appear unresolved, with angular
size $\leq 0\rlap.{''}2 - 0\rlap.{''}3$.}
\end{center}
\end{table}

\clearpage

\begin{table}
\begin{center}
\small
\caption{Parameters of the 7 mm Sources in NGC 6334 I(N).\label{tbl-3}}
\begin{tabular}{lccccc}
\tableline\tableline
 &\multicolumn{2}{c}{Position$^a$} & Total Flux
& & Estimated \\
\cline{2-3}
Source &  $\alpha$(J2000) & $\delta$(J2000) & Density (mJy) &
Deconvolved Angular Size$^b$ & Mass ($M_\odot$) \\
\tableline
SMA 6 & 17 20 54.594 & -35 45 17.01 & 3.8$\pm$1.0
& $\leq 1\rlap.{''}0$ & 4 \\
SMA 1a & 17 20 55.170 & -35 45 05.26 & 13.6$\pm$2.4
& $1\rlap.{''}9 \pm 0\rlap.{''}5 \times 1\rlap.{''}3 \pm 0\rlap.{''}4;~ +42^\circ
\pm 36^\circ$ & 4  \\
SMA 1b & 17 20 55.199 & -35 45 03.43 & 11.0$\pm$1.8
& $1\rlap.{''}4 \pm 0\rlap.{''}4 \times 1\rlap.{''}2 \pm 0\rlap.{''}5;~ +164^\circ
\pm 33^\circ$ & 3  \\
SMA 1c & 17 20 55.260 & -35 45 02.42 & 6.8$\pm$1.5
& $\leq 1\rlap.{''}0$ & 2  \\
\tableline
\end{tabular}
\tablenotetext{a}{Units of right
ascension are hours, minutes, and seconds
and units of declination are degrees, arcminutes, and arcseconds. Positional accuracy
is estimated to be $0\rlap.{''}2$.}
\tablenotetext{b}{Major axis $\times$ minor axis; position angle of major axis.}
\end{center}
\end{table}







\begin{thebibliography}{}

\bibitem[Beuther et al. (2005)]{beu05} Beuther, H., Thonwirth, S., Zhang, Q., Hunter, T. R.,
Megeath, S. T., Walsh, A. J., \& Menten, K. M. 2005, \apj, 627, 834

\bibitem[Carral et al. (2002)]{car02} Carral, P., Kurtz, S. E., Rodr\'\i guez,
L. F., Menten, K., Cant\'o, J., \& Arceo, R. 2002, \aj, 123, 2574

\bibitem[Cesaroni et al. (2006)]{ces06}
Cesaroni, R., Galli, D., Lodato, G., Walmsley, C. M. \& Zhang, Q.
2006, in Protostars and Planets V,
ed. B. Reipurth, D. Jewitt, \& K. Keili
(Tucson: Univ. Arizona Press), in press

\bibitem[Guilloteau \& Dutrey (1998)]{gui98} Guilloteau, S. \& Dutrey, A. 1998,
A\&A, 339, 467

\bibitem[Hunter et al. (2006)]{hun06} Hunter, T. R., Brogan, C. L., Megeath, S. T.,
Menten, K. M., Beuther, H., \& Thorwirth, S. 2006, submited to \apj.

\bibitem[Laughlin \& Bodenheimer (1994)]{lau94} Laughlin, G. \&
Bodenheimer, P. 1994, ApJ, 436, 335 

\bibitem[Lodato \& Rice (2005)]{lod05} Lodato, G. \& Rice, W. K. M.
2005, MNRAS, 358, 1489

\bibitem[Megeath \& Tieftrunk (1999)]{meg99} Megeath, S. T. \& Tieftrunk, A. R. 1999, \apj, 526, L113

\bibitem[Neckel 1978]{nec78} Neckel, T. 1978, A\&A, 69, 51

\bibitem[Patel et al. 2005]{pat05} Patel, N. A., Curiel, S., Sridharan, T. K.,
Zhang, Q., Hunter, T. R., Ho, Paul T. P., Torrelles, J. M.,
Moran, J. M., G\'omez, J. F., \& Anglada, G. 2005, \nat, 437, 109

\bibitem[]{re86} Reynolds, S. P. 1986, ApJ, 304, 713

\bibitem[] {ro98} Rodr\'\i guez, L. F., D'Alessio, P., Wilner, D. J., Ho, P. T. P.,
Torrelles, J. M., Curiel, S., G\'omez, Y., Lizano, S., Pedlar, A., Cant\'o, J.,
\& Raga, A. C. 1998, \nat, 395, 355

\bibitem[Sandell (2000)]{san00} Sandell, G. 2000, A\&A, 358, 242

\bibitem[Schreyer et al. (2006)]{scr06} Schreyer, S., Semenov, D., Henning, Th.,
\& Forbrich, J. 2006, \apj, 637, L129

\bibitem[Shepherd \& Kurtz (1999)]{she99} Shepherd, D. S. \& Kurtz, S. E.
1999, ApJ, 523, 690

\bibitem[Shepherd et al. (2001)]{she01} Shepherd, D. S., Claussen, M. J.,
\& Kurtz, S. E. 2001, Science, 292, 1513

\bibitem[Sodroski et al. (1997)]{sod97} Sodroski, T. J., Odegard, N., Arendt, R. G., Dwek, E.,
Weiland, J. L., Hauser, M. G., \& Kelsall, T. 1997, \apj, 480, 173 

\bibitem[Sollins \& Ho (2005)]{sol05} 
Sollins, P. K. \& Ho, P. T. P. 2005, ApJ, 630, 987

\bibitem[Thorwirth et al. (2003)]{thor03}
Thorwirth, S., Winnewisser, G., Megeath, S. T., \& Tieftrunk, A. R. 2003, 
in ASP Conf. Ser. 287, Galactic Star Formation Across the Stellar Mass Spectrum, ed. 
J. M. De Buizer \& N. S. van der Bliek (San Francisco: ASP), 257

\bibitem[Wilner et al. (2005)]{wil05} 
Wilner, D. J., D'Alessio, P., Calvet, N., Claussen, M. J., \& Hartmann, L. 2005, ApJ, 626, L109

\bibitem[Yorke et al. (1995)]{yor95} Yorke, H. W.,
Bodenheimer, P,. \& Laughlin, G. 1995, ApJ, 443, 199

\end{thebibliography}
\end{document}